\documentclass[12pt]{article}
\usepackage{amssymb,latexsym}
\usepackage{amsmath}

\columnwidth\textwidth

\begin{document}

\newtheorem{df}{Definition}
\newtheorem{thm}{Theorem}
\newtheorem{lem}{Lemma}

\begin{titlepage}

\noindent

\begin{center}
{\LARGE Intrinsic and extrinsic properties of quantum systems}

\vspace{1cm}

P. H\'{a}j\'{\i}\v{c}ek \\
Institute for Theoretical Physics \\
University of Bern \\
Sidlerstrasse 5, CH-3012 Bern, Switzerland \\
hajicek@itp.unibe.ch \\
\vspace{5mm}
and \\
\vspace{5mm}
J. Tolar\\
Department of Physics\\
Faculty of Nuclear Sciences and Physical Engineering\\
Czech Technical University\\
B\v{r}ehov\'{a} 7, CZ-11519 Prague, Czech Republic\\
jiri.tolar@fjfi.cvut.cz

\vspace{1cm}


January 21, 2008 \\


\vspace*{5mm}

\nopagebreak[4]

\begin{abstract}
The paper attempts to convince that the orthodox interpretation of
quantum mechanics does not contradict philosophical realism by
throwing light onto certain properties of quantum systems that seem
to have escaped attention as yet. The exposition starts with the
philosophical notions of realism. Then, the quantum mechanics as it
is usually taught is demoted to a mere part of the theory called
phenomenology of observations, and the common impression about its
contradiction to realism is explained. The main idea of the paper,
the physical notion of intrinsic properties, is introduced and many
examples thereof are given. It replaces the irritating dichotomy of
quantum and classical worlds by a much softer difference between
intrinsic and extrinsic properties, which concern equally
microscopic and macroscopic systems. Finally, the classicality and
the quantum measurement are analyzed and found to present some still
unsolved problems. A possible way of dealing with the
Schr\"{o}dinger cat is suggested that is based on the intrinsic
properties. A simple quantum model of one classical property
illustrates how our philosophy may work.
\end{abstract}

\end{center}

\end{titlepage}

\section{Introduction}
Quantum mechanics does not seem to be fully understood even after
about eighty years of very successful existence and there is a lot
of work being done on its interpretation today (e.g.,
\cite{Griffiths,Nikolic}). The present paper describes an approach
to its conceptual foundation from a new point od view. After some
clarification of relevant philosophical notions, it gives a short
review of quantum mechanics as it is usually understood. We propose
that this constitutes only a part of the whole theory and call it
phenomenology of observation. The other part is based on the concept
of intrinsic properties. This may be quite crucial for understanding
of quantum mechanics. It seems that it has never been explicitly
mentioned and explained, probably because those who use quantum
mechanics in their everyday work view it as obvious while those who
are engaged in philosophy have not noticed it. We give many examples
of the intrinsic properties and try to build some systematical
picture. The paper sketches the basic lines of a project dealing
with the only important unsolved problems in the conceptual
foundations, that of the origin of classical properties and that of
quantum measurement. A simple quantum model of a classical property,
the length of a solid body, is constructed in the Appendix. The
technical knowledge of quantum mechanics in the extent of, say,
\cite{Peres} will be assumed.

\section{Realism}
The {\em realism} seems to be the main apple of discord and the open or
hidden subject of most discussions on quantum mechanics (e.g.,
\cite{d'Espagnat,Isham}). Let us explain what the realism will mean in the
present paper.

Realism is an important hypothesis. It claims that Nature really exists and
is observer independent. It is not just a construct of human mind but
people are themselves a part of Nature and their thinking is based on
natural physico-chemical processes in their brain. Realism explains a lot of
coincidences in different observations or observations done by different
people that would else appear very strange. Nature is the object of study for
sciences.

To discriminate the scientific realism from its naive variant, we distinguish
our knowledge of reality from the reality itself. A very important part of our
realism hypothesis is the assent that {\em any} knowledge that we may
have about reality is incomplete and approximative. Still, it may be
successful in leading us to valid predictions within certain accuracy limits.
We do interpret this success by saying that the knowledge truly captures
some aspect of reality. From this point of view, questions such as whether
quantum mechanics is incomplete or not or whether a quantum state
describes reality or only some knowledge about it are incorrectly
formulated ones. The incompleteness of our knowledge has even a
practical, methodical feature. We usually isolate some aspect of Nature and
construct a model of it. The model can be a 'simplified' one, i.e., it may
disregard a lot of things that usually accompany the modelled aspect. Still,
it can be true in revealing a real property of Nature in the above
pragmatical sense.

\section{Phenomenology of observation}
We maintain that the ultimate aim of quantum mechanics is to study real
properties of real quantum systems.

The word 'property' is introduced here to have a general notion of
observable properties concerning quantum systems. For instance, the
values of the quantities that are called {\em observables} in
quantum mechanics are properties. Our main idea is that the values
of the observables form only a subset of properties of quantum
systems. Let us call these properties {\em extrinsic}\footnote{More
generally, extrinsic properties can be described as linear subspaces
in the Hilbert space of the system. They represent the mathematical
counterpart of the so-called YES-NO experiments \cite{Piron}. The
set of linear subspaces admits the usual operations on conjunction
(linear hull), disjunction (intersection) and negation (othogonal
complement), but the resulting orthocomplemented lattice is not a
Boolean lattice \cite{BvN}. As it is well known, the set of
'classical' properties of a single system forms a Boolean lattice
(of subsets of classical phase space). If we pretend that the
extrinsic properties of a quantum system are properties of a
well-defined single system, then we are lead to abandon the ordinary
logic and introduce the so-called {\em quantum logic}. But this
pretention is against all logic because the extrinsic properties are
properties of many different systems each consisting of the quantum
system plus some apparatus.}.

Quantum mechanics is usually understood as an abstract theory of the
extrinsic properties, consisting of the usual stuff about Hilbert
spaces, states and observables. The theory is abstract in the sense
that it does not work with any specific system. In most
presentations of quantum mechanics, the greatest attention is
dedicated to this part so that a wrong impression arises that
quantum mechanics does not contain anything else. The consequence of
the impression are many utterances such as Bohr's, "There is no
quantum concept." We consider this to be grossly exaggerated or even
wrong. To prevent such confusion, let us call the abstract theory of
the extrinsic properties the {\em phenomenology of observation}.

The existence of the phenomenology is a conspicuous feature because in
none of the older physical theories do similar parts play such a
fundamental role as it does in quantum mechanics. The subjectivistic,
operationalistic or positivistic flavour of this part of quantum mechanics is
of course due to its being a theory of human observation and does not
imply anything like non existence of observer-independent reality.

The crucial point of the phenomenology is the existence of {\em classical
systems}. These are arrangements of bodies and fields to which classical
mechanics, electrodynamics and thermodynamics are applicable as very
good approximations. Moreover, to describe the properties of these systems
that are relevant for the quantum observations, quantum mechanics itself
is not needed. Their corresponding classical properties are directly
observable and amenable to manipulations by people. Specifying and
bringing into being classical properties is possible for us so that it can be
said that we control the classical conditions of the experiment.

At the beginning of any quantum experiment or observation stands
what is usually called a {\em preparation}. The name is somewhat
misleading. What is meant is a set of classical conditions which the
quantum system to be observed is subject to before the observation.
This can, but need not, include some human activity in laboratory.
For example, we can know that a quantum system inside the Sun is the
plasma with a given composition and that its classical conditions
are certain temperature and pressure. Sufficiently precise
description of the classical conditions must be given so that the
same preparation is in principle reproducible. Thus, a series of
repeated experiments is possible, and the set of quantum systems
obtained by repeating the experiment is called {\em ensemble}.
Clearly, the notion of ensemble is in many aspects closely connected
to that of preparation.

With a specific preparation, a state of the quantum system is
associated. More precisely, if we repeat the experiment so that all
classical conditions remain the same, the quantum state is always
the same by definition. The state is mathematically described by a
{\em state operator} in the Hilbert space. In this sense, the state
generally represents our knowledge on the system. This knowledge can
have different degrees of certainty, that is, different entropies.
Maximal certainty with entropy zero is represented by a projector to
a one-dimensional subspace of the Hilbert space and the state is
called {\em pure}. The minimal certainty state is proportional to
unity and the state is called {\em completely chaotic}. Better, it
expresses our complete ignorance about the classical conditions. Its
entropy is $\ln N$, where $N$ is the dimension of the Hilbert space.
Even the completely chaotic state of a given system does still
contain non-trivial information, namely that about some intrinsic
properties of the system.

At the end of any quantum experiment there is what is often called a {\em
registration}. It is an interaction of a individual quantum system in a
specific state with a classical system, the {\em measuring apparatus}.
Ideally, each measuring apparatus is mathematically represented by an
observable, a self-adjoint operator in the Hilbert space of the system.

We are free to choose an observable from the set determined by the
structure of the system and carry out the corresponding registration
on any state $\rho$ that has been prepared. These choices form the
first set of alternative possibilities. The measurement leads to the
observable acquiring a definite value. All possible values of the
observable that can be obtained are the eigenvalues of the
corresponding operator; they form the second set of alternative
possibilities. These cannot however be chosen freely and we know
only the probabilities of these possible results. That is, if the
state $\rho$ is prepared many times, then the same measurement will
in general not give the same result each time. The probabilities are
mathematically determined by the state operator together with the
operator of the observable. Thus, a state contains many different
kinds of information.

Generally, the values of observables do not directly refer to the quantum
system alone but to the composite system of the quantum and classical
systems in interaction. As such, even they are real (observer independent):
they are the 'beables' of John Bell \cite{bell1}. The idea that they refer to
the composite system and not the quantum system alone suggests why the
information about results of measurements need not exist before the
measurements.

Thus, the phenomenology of observation describes directly only
processes and properties concerning classical bodies; it is even not
necessary to assume that any quantum systems exist. There is nothing
mysterious about this. We cannot observe a quantum system directly.
We have to use the classical traces that the quantum systems leave
on classical systems which they are interacting with. Moreover, the
{\em classicality} of the macroscopic bodies is crucial for the
statistical interpretation of quantum mechanics. The classicality of
a measuring apparatus means among others that it yields a definite
value for each individual measurement and that all possible values
form mutually exclusive alternatives. Only then it is sensible to
speak about probabilities.

It is a miracle that a systematic and beautiful mathematical theory
exists describing these phenomena. In fact, the phenomenology has
been formulated in a rigorous mathematical way by Ludwig
\cite{ludwig} and by Kraus \cite{kraus} and has evolved into a
broadly used theory today.

\section{The intrinsic properties of quantum systems}
Our point of view is that there are properties of quantum systems that are
not quantum-mechanical observables. They can be ascribed directly to
quantum systems and assumed to be real (observer independent) without
the danger of paradoxes. That's why we call these {\em intrinsic}
properties.

The first among the intrinsic properties is the {\em structure of a
quantum system}. Quantum mechanics contains well-defined rules about
what can such a structure be. For example, in the non relativistic
case, there must be a definite number\footnote{There are
non-relativistic systems, in which some particle numbers are
variable, such as those of quasi-particles in solid state physics.
Of course, these particle numbers do not belong to the structure of
the systems and they are not intrinsic properties but extrinsic
properties in our conception.} of some particles with definite
masses and spins. The particles interact with each other by a
definite potential function. There are important further rules about
symmetries, etc. For a relativistic case, there are analogous rules:
we have fields of certain (bare) masses and spins interacting by
means of suitable interaction Lagrangians involving (bare) coupling
constants.

In the previous paragraph, we have distinguished non-relativistic and
relativistic systems not only to avoid the problem of how the
non-relativistic systems are to be defined as some special cases
(approximations) of the relativistic ones. More important reason has been
to show that the difference between intrinsic and extrinsic properties can
be model dependent. The model itself, in turn, is constructed according to
the situation to be considered. What is relevant is that every quantum
model exhibits each of the two kinds of properties, both intrisic and
extrinsic.

For example, the model of hydrogen atom consists of two particles, proton
and electron, that have certain masses and spins. These constituents
interact with each other by means of the Coulomb potential that is
determined by their charges.

The next set of rules allows us to determine the quantum observables that
can be measured on the system. For example, each particle contributes to
the observables by three coordinates and three momenta. Thus, in the
hydrogen case, there will be (in addition to other observables) six
coordinates and six components of momenta. The {\em set of observables
that can be measured on a given system} is its intrinsic property and this
information is different from that about the values of these
 observables\footnote{More precisely, the set of observables can be
 embedded in
a structure of the so-called $C^*$-algebra that represents a part of
the physical structure of the system \cite{thirring}. Thus, it is an
intrinsic property of it. Moreover, such algebras have a
representation on a Hilbert space---the Hilbert space of the system.
Of course, for systems with finite number of degrees of freedom, the
Hilbert space representation is uniquely defined (up to unitary
equivalence) by the algebra, so it does not contain any further
information on an independent structure of the system, but the
algebras of relativistic fields possess many inequivalent
representations of which only few are physical, corresponding to
different phases of the system. A physical representation is clearly
an independent intrinsic property of the field.}. An important point
is that many intrinsic properties are not accessible by direct
measurements but are only determined via measurement of the
extrinsic ones. We can see here also that the theories of the
intrinsic and extrinsic properties cannot be separated from each
other.

The structure and the observables of a system are used to set up the
Hamiltonian of non-relativistic or the action functional for
relativistic systems according to further basic rules. The {\em form
of the Hamiltonian or the action} are mathematical expressions of
the structure and thus intrinsic properties. Using the Hamiltonian
or the action, we can write down the dynamical laws---the
Schr\"{o}dinger equation or the path-integral formula---from which
other important intrinsic properties can be calculated, for example
the spectrum of the hydrogen atom. The spectrum is clearly an
intrinsic property of the hydrogen atom that can be ascribed to the
system itself independently of any measurement. This will not lead
to any contradictions with other measurements or ideas of quantum
mechanics. We can recognize the system with the help of its
intrinsic properties. For example, if we detect light from somewhere
deep in the Universe and find the Balmer series in its spectrum,
then we know that there is hydrogen there. The numbers such as cross
sections, branching ratios etc. are further examples of intrinsic
properties

These rules form a part of basic principles of quantum mechanics. It is
important to realize that they are not directly derivable from evidence; they
are the basic hypotheses of the theory. The role of these principles is to
define a specific class of models for quantum systems. For each system, we
can attempt different possible models, calculate the extrinsic properties of
each and compare with the experimental evidence gained in a number of
quantum experiments. In this way, the models can be confirmed or
disproved.

Thus, the intrinsic properties of a given quantum system are assumed
to exist independently of observers or observations. Still, such an
observer-independent information about an individual quantum system
is not complete in the sense that intrinsic properties do not
determine everything that can be ascertained about it; the extrinsic
properties that had to be added are observer dependent, or better,
measurement dependent. However, this seems to be a general property
of modern physics that does not contradict realism.

For instance in the special relativity, there are invariants that are
independent of the choice of inertial frames, which, in turn, depend on, or
represent observers. More precisely, a relativistic system defines a set of
invariants that are formed from the variables of the system alone
('intrinsic' invariants). For example, such variables are the components
$P^\mu$ of the four-momentum of a free particle ${\mathcal S}$ and the
Poincar\'{e} invariant is the squared length $P^\mu P_\mu$. Then, there
is another set of invariants describing relations of the system with other
systems ('extrinsic' invariants) and which can replace the inertial frame
coordinates or components. For example, one can consider the inertial
frame as a physical system ${\mathcal F}$ consisting of a radar station,
three gyroscopes and a clock; then the coordinates of the particle with
respect to this frame can be calculated as Poincar\'{e} invariants formed
from the variables of the two systems ${\mathcal S}$ and ${\mathcal F}$.
For a complete description of a relativistic system, the intrinsic invariants
do not suffice and coordinates with respect to a frame (or extrinsic
invariants) are needed.

Another important point to keep in mind is the {\em dependence of
the notion of intrinsic property on the notion of quantum sytem}.
The existence of a hypothetical quantum system need not itself be,
under some conditions, an intrinsic property. As an example,
consider a quantum system consisting of an electron and a proton.
Let us prepare the electron and the proton in pure states that are
spatially well separated and let the total energy in their
center-of-mass frame be lower than the binding energy of hydrogen
atoms. Such a system is not a hydrogen atom, because a hydrogen atom
in this energy state would be bound. Now suppose that our initial
state evolves. Then, there is a probability that the electron will
be captured and a bound state will form. In general, the final state
after the evolution will be a linear superposition of states, some
of them representing the electron and proton well separated (no
hydrogen atom), others being states of a (bound) hydrogen atom.
Thus, the existence of hydrogen atom is not an intrinsic property of
our original quantum system. In general, it follows that the
existence of composite quantum systems is relative and
approximative: under some conditions (that is what is `relative`),
the assumption of the existence of a quantum system can be
successful in giving valid predictions in certain accuracy limits
(that is what is `approximative`).

Now, let us turn to the macroscopic, 'classical', world. This paper is going
to propose that
\begin{quote}{\em all variables describing classical (i.e., geometrical,
mechanical and thermodynamical) state of macroscopic systems can be
obtained from quantum mechanics as intrinsic properties}.
\end{quote}
Indeed, they can be ascribed to the systems themselves and the
assumption that they exist independently of observation does not lead to
any paradoxes. The (implicit) idea that at least some properties of
macroscopic systems are their intrinsic properties that can be calculated
by the usual methods of quantum mechanics has been very fruitful in the
past. For example, the solid state physics explains the rich physical
properties of solids (such as electrical conductivity), which are of course
intrinsic.

Everything what we can measure on classical systems has a form of
average value (we adopt this point of view, which is originally due
to Exner \cite{Exner}, p.~669, and Born \cite{Born}) and its
dispersion (mean quadratic deviation). Such a property is associated
with a whole ensemble of systems rather than with an individual one.
We can, however, generalize the notion of systems to include such
ensembles. They are defined by the conditions under which the
individual systems are accepted as their elements. Thus, average
values can be considered as intrinsic properties, too. In many
cases, the associated dispersions are small in comparison with the
average values themselves. Then, one usually speaks of values that
concern the individual systems of the ensemble. It seems that all
values in the classical physics that are pertinent to individual
systems have this character.

An interesting question is that about the origin of the dispersion of
classical quantities. It is often assumed that improvements in measuring
techniques will in principle, in some limit, lead to zero dispersion. This is in
agreement with the classical theory such as mechanics. It predicts that the
trajectories are completely sharp if the initial data are so, and does not put
any limit on the accuracy with which the initial state can be prepared. The
point of view adopted here is different: some part of the dispersions cannot
ever be removed and the classical theories are only approximative models.

The macroscopic systems are highly complicated from the quantum
point of view. It seems that this may be a source of intrinsic
properties that do not make sense for small quantum systems. Indeed,
an example is provided by molecules of the deoxyribonucleic acid.
Their structures become richer with their length, their number grows
(roughly) exponentially with the number of the four constituents
because all the possible orderings of the constituents define
different structures. The rich intrinsic properties of large systems
might also enable a new approach to quantum cosmology without the
'wave function of the Universe'.

Another kind of intrinsic properties possessed exclusively by
macroscopic systems are the thermodynamical ones. They include
average values (expectation values, mean values) of several
quantum-mechanical observables, such as energies or particle numbers
of subsystems. For macroscopic subsystems, these observables form,
on the one hand, only a very small subset of the whole observable
algebra of the system and, on the other, define its macroscopic
state by their average values. Moreover, they have negligible
relative dispersions (variances, mean square deviations) in states
that are close to thermodynamic equilibrium. The existence of the
equilibrium state and the fact that it evolves spontaneously from
overwhelmingly large set of initial microstates is a further
intrinsic property of the macroscopic systems.

Thermodynamic properties can to a large extent be derived from quantum
mechanics \cite{thirring} in the limit of infinite particle number. The
equilibrium state is clearly compatible with very many microscopic
quantum states and so it gives a very incomplete information from the
quantum point of view. The equilibrium can be defined as a maximum
entropy\footnote{The term 'entropy' always means the von Neumann
entropy in this paper.} state under specific macroscopic conditions. For
example, the Gibbs state is defined by the maximum of entropy at a given
average energy. This property makes it to a very good approximation to
what can mostly be observed. A simplified model showing in some detail
how these ideas on macroscopic systems could work is presented in the
Appendix.

To summarize, quantum mechanics comprises the knowledge gained by
long experience and concerning the possible structures of quantum
systems. Using this knowledge, we can construct quantum models of
newly observed systems. It further specifies how the ideas about the
structure can be used to write down quantitative laws relating
different intrinsic properties. Finally, it determines what are the
extrinsic properties of the system that can be measured. It also
includes the specific structures of the menagerie of known quantum
systems: nuclei, atoms, molecules, solid bodies, relativistic field
systems and many more.

\section{Classical properties and the quantum theory of measurement}
Our general philosophy will work satisfactorily only if the classical systems
can be considered as some special kind of quantum systems and their
classical properties can be derived from quantum mechanics. This is
known as the hypothesis of {\em universality of quantum mechanics}.

To begin the discussion, let us consider the so-called semiclassical (or
WKB) approximation. This is based on the observation that, for a number
of systems, the dynamics of the average values of a number of quantum
observables follows classical (say, Newton mechanics) trajectories. This is
surely a good start because, as we have seen, such average values can be
considered as intrinsic properties. However, the classical systems do
possess an additional crucial property: each observation of a classical
quantity gives approximately the same value equal to the average one.
Thus, we also need a negligible dispersion of these observables in most
states of macroscopical systems that can be observed.

In quantum mechanics, a general method to construct states with
large dispersion is provided by the superposition principle. Indeed,
if my chair were a quantum system, then its average position can of
course be in my room, but nothing seems to prevent it to be in a
linear superposition of states each of them representing a position
in another room. Such a chair cannot be considered to be in any of
the rooms; rather, it is in all of them simultaneously. This
interpretation of the linear superposition follows from observations
such as the well-known two-slit experiment. Then, even if the
average position of the chair is correct, the mean quadratic
deviation is of the order of the apartment size. Long ago,
Schr\"{o}dinger invented a paradox that is well-known under the name
´Schr\"{o}dinger cat´ to help visualising the problem.

It turns out that what one needs is the validity of {\em macroscopic
realism} \cite{Leggett} and we are faced with the problem to derive
it from quantum mechanics. It has been defined as follows.
\begin{enumerate}
\item A macroscopic object which has available to it two or more
macroscopically distinct states is at any given time in a definite one of
those states.
\item It is possible in principle to determine which of these states the
system is in without any effect on the state itself or on the subsequent
system dynamics.
\item The properties of ensembles are determined exclusively by initial
conditions (and in particular not by final conditions).
\end{enumerate}

This property of macroscopic systems is necessary for the quantum
phenomenology of observation itself to make sense. Indeed, suppose
that a measurement is being done. Suppose that the apparatus is a
quantum system and that it is in a linear superposition of different
eigenstates of its pointer after the measurement. This superposition
state is physically different from a proper mixture of states, each
being a definite eigenstate of the pointer. Hence, we cannot just
read off the unique value that is to result in each case of the
repeated measurement. To bring the apparatus into such a proper
mixture state, we had to make an additional measurement on the
apparatus by an additional apparatus. And so on.

The usual models of measuring apparatus \cite{JvN,Jauch} assume that
the states of its pointer are eigenstates of some operator---that is, they are
extrinsic properties of the apparatus. Then, what one had to achieve is that
the state of the apparatus after a measurement would be a proper mixture
of the pointer states. The property of a state operator to be a proper
mixture and the set of states that form the corresponding ´real´ alternatives
can be e.g.\ created during the preparation or it follows from some kind of
superselection rules that forbid linear superposition of the pointer states.
The information about such properties is never contained in the form of the
state operator itself.\footnote{This seems to contradict the claim (that can
be found in any textbook) that the state operator contains all information
about measurable properties of the state. The contradiction is only
apparent because the word ´measurable´ used by the claim means
´obtainable through quantum measurement performed on the prepared
state.´ What we need is, however, a property existing before such
measurement. More about our interpretation of state operators, see
\cite{TH}.} For example, if a state operator is diagonalized in an
orthonormal basis of states, then it does not follow that it is a proper
mixture of the states. Such an assumption leads to contradictions: the
simplest counterexample is a proper mixture of two non-orthogonal states.

There is much activity in this field. Let us mention the quantum
decoherence theory \cite{zurek, zeh}, the Coleman-Hepp theory
\cite{Hepp,Bell3,Bona} and its modifications \cite{Sewell} and
theories based on some coarse graining \cite{Peres,poulin,Kofler}.
At the present time, the above problem does not seem to be solved in
a completely satisfactory way, see also \cite{d'Espagnat, bell2}.

The difficulties may come from having a wrong quantum model of the
measurement process. The essential feature of this model is to view
the classical properties of macroscopic quantum systems as their
extrinsic properties. Thus, a proposal seems to be natural that the
model of the measuring process must be modified, so that the
classical properties of macroscopic quantum systems will be their
intrinsic properties. In fact, this follows directly from the
principle of macroscopic realism.

To summarize: We have found that the current quantum models of
macroscopic body, of the quantum measurement and of the measuring
apparatus are not completely satisfactory. Indeed, a real apparatus
does yield definite values of measured observables while the quantum
model of the apparatus fails to do so. Of course, the problem does
not prevent us from using quantum mechanics successfully. We have
been always having the provisional way of how to do it, provided by
the old Born formula and the splitting of the world into its
classical and quantum parts. However, our theory of intrinsic
properties needs a solution to this problem and can help to obtain
one. More work is necessary.

\appendix
\section{Quantum model of classical body}
We are going to construct a simplified quantum model of an ordinary
classical body. In accordance with our previously stated project, we ought
to obtain all of its ordinary classical properties, geometrical, mechanical
and thermodynamical, as intrinsic properties of the corresponding
quantum system. This entails that, first, the quantum structure of the
system must be defined, second, the basic intrinsic properties such as the
spectrum calculated, and, third, further intrinsic properties derived. As yet,
we can define, e.g., the length of the body.

\subsection{The structure and the Hamiltonian}
We shall consider a linear chain of $N$ identical particles of mass
$\mu$ distributed along the $x$-axis with the Hamiltonian
$$
  H = \frac{1}{2\mu}\sum_{n=1}^N p_n^2 +
\frac{\kappa^2}{2}\sum_{n=2}^N (x_n - x_{n-1} - \xi)^2,
$$
involving only nearest neighbour elastic forces. Here $x_n$ is the
position, $p_n$ the momentum of the $n$-th particle, $\kappa$ the
oscillator strength and $\xi$ the equilibrium interparticle
distance. The parameters $\mu$, $\kappa$ and $\xi$ are intrinsic
properties (the last two defining the potential function).

This kind of chain seems to be different from most that are studied
in literature: the positions of the chain particles are dynamical
variables so that the chain can move as a whole. However, the chain
can still be solved by methods that are described in
\cite{Kittel,Rutherford}.

\subsection{The modes}
After the transformation
\begin{equation}\label{xy}
  x_n = y_n + \left(n - \frac{N+1}{2}\right)\xi,
\end{equation}
the potential becomes a quadratic form
$$
  V = \frac{\kappa^2}{2}\sum_{n=2}^N (y_n - y_{n-1})^2.
$$
and the equations of motion read
\begin{eqnarray*}
\mu\ddot{y}_n &=& \kappa^2(y_{n+1}-2y_n+y_{n-1})\quad\forall\quad
1<n<N, \\
\mu\ddot{y}_1 &=& \kappa^2(y_2-y_1), \\
\mu\ddot{y}_N &=& \kappa^2(-y_N+y_{N-1}).
\end{eqnarray*}
To simplify the equations, we add fictitious points $0$ and $N+1$ to
the chain and require the additional variables $y_0$ and $y_{N+1}$
to satisfy the boundary conditions of free ends,
$$
y_0 = y_1,\quad y_{N+1} = y_N.
$$
Then, the equations of motion can be written as
$$
\mu\ddot{y}_n = \kappa^2(y_{n+1}-2y_n+y_{n-1})\quad\forall\quad 1\leq
n\leq N.
$$
 By the standard method of modes, we substitute $ y_n = Y_n\cos\omega t$
and obtain the linear system for the mode amplitudes $Y_n$,
\begin{equation}\label{algeqs}
Y_{n+1} + Y_{n-1} = \left(2-\frac{\mu}{\kappa^2}\omega^2\right)Y_n,
\quad\forall\quad 1\leq n\leq N,
\end{equation}
with boundary conditions \cite{Rutherford}
\begin{equation}\label{BC}
Y_0 = Y_1,\quad Y_{N+1} = Y_N.
\end{equation}

Some general properties of this system can be obtained as follows.
Introducing the shorthand notation
$$
V_{ij} := \frac{1}{2}\frac{\partial^2V}{\partial y_i\partial
y_j}(0,\cdots,0),
$$
the equations of motion and the linear system take the form
$$
\ddot{y}_i = -\frac{\kappa^2}{\mu}\sum_{j=1}^NV_{ij}y_j, \qquad
\sum_{j=1}^NV_{ij}Y_j = \frac{\mu\omega^2}{\kappa^2}Y_i.
$$
Hence, there must be $N$ modes with amplitudes $\{Y_i\}$ that
diagonalize the symmetric matrix $V_{ij}$ and they can be chosen to
be orthonormal with respect to the scalar product $\sum_{j=1}^N
Y_jY'_j$.

We can observe further that the system (\ref{algeqs}), (\ref{BC}) is
invariant with respect to the inversion of the chain order,
$$
Y'_n = Y_{N+1-n},
$$
so that the modes can be separated into even and odd ones. The next
step are the harmonic solutions of (\ref{algeqs}), (\ref{BC}): for
even modes,
\begin{equation}\label{evenm}
Y_n = A^+(k)\cos\left[kn-\frac{k(N+1)}{2}\right],
\end{equation}
and for the odd ones,
\begin{equation}\label{oddm}
Y_n = A^-(k)\sin\left[kn-\frac{k(N+1)}{2}\right],
\end{equation}
where $A^\pm(k)$ are normalization factors. In both cases, we obtain
the dispersion relation
\begin{equation}\label{dispersion}
\omega(k) = \frac{2\kappa}{\sqrt{\mu}}\sin\frac{k}{2}.
\end{equation}

From the two boundary conditions, only one is now independent. For the
even modes, equation $Y_0 = Y_1$ becomes
$$
\cos\left[\frac{k(N+1)}{2}\right] = \cos\left[\frac{k(N-1)}{2}\right],
$$
which is equivalent to
$$
\sin\frac{kN}{2}\sin\frac{k}{2} = 0 \quad \Leftrightarrow \quad
  k = \frac{2m}{N}\pi,
$$
where $m$ is any integer. Similarly, for the odd modes we obtain
$$
k = \frac{2m-1}{N}\pi.
$$
Altogether there are $N$ modes: we obtain finally, for each $N$,
\begin{equation}\label{rangem}
k_m = \frac{m}{N}\pi, \qquad m = 0,1,\cdots,N-1,
\end{equation}
 and
\begin{equation}\label{spectr}
\omega_m = \omega(k_m) =
\frac{2\kappa}{\sqrt{\mu}}\sin\frac{m}{N}\frac{\pi}{2},\quad
\end{equation}
where even (odd) $m$'s correspond to the even (odd) modes and
Eqs.~(\ref{evenm}) ((\ref{oddm})) must be used for the $Y$'s. We can
see that the spectrum is non-degenerate and lies in the interval
 $\omega\in [0,2\kappa/\sqrt{\mu})$. The normalization factors
 $A^\pm(m)$ are obtained easily using Eq.~(\ref{rangem}):
for any $N$ and $m=0$
\begin{equation}\label{factor0}
A^+(0) = \frac{1}{\sqrt{N}};
\end{equation}
for $m = 1,2,\dots,N-1$ a longer calculation gives
\begin{equation}\label{factorm}
 [A^\pm(k_m)]^{-2} = \frac{N}{2} \pm \frac{1}{2}
 \frac{\sin m\pi}{\sin \frac{m\pi}{N}}=\frac{N}{2}, \quad
 \mbox{i.e.,} \quad
A^\pm(k_m) = \sqrt{\frac{2}{N}}.
\end{equation}

The results that have been obtained can be used to transform the
Hamiltonian to a diagonal form. Let us denote the mode amplitudes
that correspond to the parameter value $m$ by $Y^m_n$. Then, we can
transform the original variables $y_n$ and $p_n$ to normal mode
variables $u_m$ and $q_m$,
\begin{equation}\label{yu}
y_n = \sum_{m=0}^{N-1}Y^m_nu_m, \qquad p_n =
\sum_{m=0}^{N-1}Y^m_nq_m.
\end{equation}
 As the transformation of both
$y$'s and $p$'s is orthogonal, the new variables are canonically
conjugate and the Hamiltonian becomes
$$
 H = \frac{1}{2\mu}\sum_{m=0}^{N-1}q_m^2 + \frac{\mu}{2}\sum_{m=0}^{N-
1}\omega_m^2u_m^2.
$$

Consider the terms with $m=0$. We have $k_0=0$, $\omega_0=0$,  and
$Y^0_n=1/\sqrt{N}$. Hence,
$$
u_0 = \sum_{n=1}^N\frac{1}{\sqrt{N}}y_n, \quad q_0 =
\sum_{n=1}^N\frac{1}{\sqrt{N}}p_n,
$$
so that
$$
u_0 = \sqrt{N}X,\quad q_0 = \frac{1}{\sqrt{N}}P,
$$
where $X$ is the center-of-mass coordinate of the chain and $P$ is
its total momentum. The 'zero' terms in the Hamiltonian then reduce
to
$$
\frac{1}{2M}P^2
$$
with $M = N\mu$ being the total mass. Thus, the 'zero mode'
describes a straight, uniform motion of the chain as a whole. The
other modes are 'phonons' with eigenfrequencies $\omega_m$, $m =
1,2,\dots,N-1$. The phonon excitation energy spectrum of the body is
built from the eigenfrequencies by the formula
\begin{equation}\label{phonons}
E = \sum_{m=1}^{N-1}\nu_m \hbar\omega_m,
 \end{equation}
where $\{\nu_m\}$ is an $(N-1)$-tuple of non-negative
integers---phonon occupation numbers.

\subsection{Numerical values}
Here, we choose the order of magnitude of the parameters to mimick
real bodies. The distances of neighbouring atoms typically are
$$
  \xi \approx 5.10^{-10}\ m
$$
to be compared with atomic radii of the order $2.10^{-10}\ m$
\cite{Kittel} or with the Bohr radius
$$
a_0 \approx \frac{\hbar}{m_{e}c\alpha} \approx 5\times 10^{-11}\,m,
$$
where $\hbar$ is the Planck constant, $m_e$ the electron mass, $c$
the speed of light and $\alpha$ the fine structure constant.

The dispersion relation \eqref{dispersion} can be estimated from the
neutron scattering measurement: e.g., for $Na$ at $90 \ K$
\cite{Kittel} the maximal frequency was found to be of the order of
$5 \ THz$. Thus, $\omega_{max} \approx 2\pi . 5.10^{12}\,Hz$, and
 $$
\omega_m = \omega_{max} \sin\left(\frac{m\pi}{2N}\right) =
\frac{2\kappa}{\sqrt{\mu}}\,\sin\left(\frac{m\pi}{2N}\right) \approx
3.10^{13}\sin\left(\frac{m\pi}{2N}\right)\,s^{-1};
$$
the corresponding energies are
$$
\hbar\omega_m \approx 3.10^{-21}\sin\left(\frac{m\pi}{2N}\right)\ J
\approx 2.10^{-2}\sin\left(\frac{m\pi}{2N}\right)\ eV.
$$
Next, let us assume that the body is in the thermodynamical
equilibrium at about $300 \ K$. Then,
 $k_BT \approx 5.10^{-21}\,J \approx 2.10^{-2}\ eV$
which corresponds to $\sin\frac{m\pi}{2N} \approx 1$.

Note that a rough estimate of the force acting on an atom displaced
from its equilibrium position in the body can also be obtained from
the known compressibility \cite{Kittel}, leading to the same order
of oscillation frequency, e.g. $4,8 \ THz$ for copper at room
temperature.

\subsection{The length of the body}
Classical properties that can be defined and calculated in our
quantum model are the average length of the body and the
corresponding dispersion. Let us define the length operator by
\begin{equation}\label{length}
   L = x_N - x_1.
\end{equation}
It can be expressed in terms of normal coordinates $u_m$ using
Eqs.~(\ref{xy}), (\ref{yu}),
$$
L = (N-1)\xi + \sum_{m=0}^{N-1}(Y^m_N-Y^m_1)u_m.
$$
The differences on the right-hand side are non-zero only for odd
values of $m$, and equal then to $-2Y^m_1$. We easily find, using
Eqs.~(\ref{oddm}), (\ref{rangem}) and (\ref{factorm}):
\begin{equation}\label{L}
L = (N-1)\xi - \sqrt{\frac{8}{N}}\ \sum_{m=1}^{[N/2]}(-
1)^m\cos\left(\frac{2m-1}{N}\frac{\pi}{2}\right)\,u_{2m-1}.
\end{equation}

The phonons of one species are excitation levels of a harmonic
oscillator, so we have
$$
u_m = \sqrt{\frac{\hbar}{2\mu\omega_m}}(a_m + a^\dagger_m),
$$
where $a_m$ is the annihilation operator for the $m$-th species. The
diagonal matrix elements between the energy eigenstates
$\mid\nu_m\rangle$ that we shall need then are
\begin{equation}\label{averu}
\langle\nu_m\mid u_m\mid\nu_m\rangle = 0,\quad \langle\nu_m\mid
u^2_m\mid\nu_m\rangle = \frac{\hbar}{2\mu\omega_m}(2\nu_m + 1).
\end{equation}

We assume that the phonons of each species form statistically
independent subsystems, hence the average of an operator concerning
only one species in the Gibbs state of the total system equals the
average in the Gibbs state for the one species. Such a Gibbs state
operator for the $m$-th species has the form
$$
\rho_m = \sum_{\nu_m=0}^{\infty}\mid\nu_m\rangle
p_{\nu_m}^{(m)}\langle\nu_m\mid,
$$
where
$$
 p_{\nu_m}^{(m)} =
Z^{-1}_{m}\exp\left(-\frac{\hbar\omega_m}{k_BT}\nu_m\right)
$$
and $Z_m$ is the partition function for the $m$-th species
\begin{equation}\label{partf}
Z_{m}(\beta) = \sum_{\nu_{m}=0}^\infty e^{-\beta\hbar\omega_m\nu_m}
= \frac{1}{1-e^{-\beta\hbar\omega_m}},
\end{equation}
where $\beta = 1/k_BT$. The thermodynamic average value of $\nu_m$
is then given by
$$
\langle\nu_m\rangle_T =
-\frac{1}{\hbar\omega_m}\left(\frac{1}{Z_m}\frac{\partial
Z_m}{\partial\beta}\right)_{\beta= (k_BT)^{-1}}
$$
and Eq.~(\ref{partf}) yields
\begin{equation}\label{avernu}
\langle\nu_m\rangle_T =
 \frac{1}{\exp\left(\frac{\hbar\omega_m}{k_BT}\right)-1}.
\end{equation}
Returning to Eq.~(\ref{L}), the average length is obtained using
(\ref{averu}),
\begin{equation}\label{averL}
\langle L\rangle_T = (N-1)\xi.
\end{equation}

Now the measure of thermodynamic fluctuations of quantity $L$ is
$$
\frac{\Delta L}{\langle L\rangle_T}=
 \frac{\sqrt{\langle L^2\rangle_T -
 \langle L\rangle_T^2}}{\langle L\rangle_T}.
$$
To estimate the dispersion $\Delta L$ to leading order for large
$N$, we start with
$$
\langle L^2\rangle_T = (N-1)^2\xi^2 +
 \frac{8}{N}\sum_{m=1}^{[N/2]}\sum_{n=1}^{[N/2]}(-1)^{m+n}
 \cos\left(\frac{2m-1}{N}\frac{\pi}{2}\right)
  \cos\left(\frac{2n-1}{N}\frac{\pi}{2}\right)
 \langle u_{2m-1}u_{2n-1}\rangle_T.
$$
Since
$$
\langle u_{2m-1}u_{2n-1}\rangle_T=\delta_{mn}\langle
u_{2m-1}^2\rangle_T,
$$
 the above formula leads to
$$
\langle L^2\rangle_T - \langle L\rangle_T^2 =
\frac{8}{N}\sum_{m=1}^{[N/2]}
 \cos^2\left(\frac{2m-1}{N}\frac{\pi}{2}\right)
 \langle u_{2m-1}^2\rangle_T,
$$
where
$$
\langle u_{2m-1}^2\rangle_T = \frac{1}{Z_{2m-1}}
 \sum_{\nu_{2m-1}=0}^\infty
 \frac{\hbar}{2\mu\omega_{2m-1}}(2\nu_{2m-1}+1)
 \exp(-\beta\hbar\omega_{2m-1}\nu_{2m-1}).
$$
Introducing dimensionless quantities
$$
x_m = \sin\left(\frac{2m-1}{N}\frac{\pi}{2}\right),\quad \gamma =
\frac{2\hbar\kappa}{k_BT\sqrt{\mu}},
$$
we can substitute $\omega_{2m-1}=(2 \kappa/\sqrt{\mu})x_m$ and
obtain the intermediate result
$$
\langle L^2\rangle_T - \langle L\rangle_T^2 =
\frac{2}{N}\frac{\hbar}{\kappa\sqrt{\mu}}
 \sum_{m=1}^{[N/2]}\frac{1-x_m^2}{x_m}
\frac{1+e^{-\gamma x_m}}{1-e^{-\gamma x_m}}.
$$
In order to extract the leading term for large $N$, we note that
$$
x_m-x_{m-1} = \frac{\pi}{N}\cos\frac{2m-1}{N}\frac{\pi}{2} +
O(N^{-2}).
$$
Then we can write
$$
\langle L^2\rangle_T - \langle L\rangle_T^2 \approx
\frac{2}{\pi}\frac{\hbar}{\kappa\sqrt{\mu}}\sum_{m=1}^{[N/2]}(x_m-x_{m-
1})f(x_m),
$$
where
$$
 f(x)=
\frac{\sqrt{1-x^2}}{x}\ \frac{1+e^{-\gamma x}}{1-e^{-\gamma x}}.
$$
 By inspection, $f$ is a decreasing function od $x$ in
the interval $(0,1)$ diverging to plus infinity at $x\rightarrow 0+$
and going through zero at $x=1$. The leading term at $x\rightarrow
0+$ is
$$
f(x) = \frac{2}{\gamma x^2}[1+O(x)].
$$
The block diagram of the sum now shows that
$$
\sum_{m=1}^{[N/2]}(x_m-x_{m-1})f(x_m) < 2x_1f(x_1) +
\int_{x_1}^1dx\,f(x).
$$
The dependence of the integral on its lower bound can be approximated by
$$
\int_{x_1}^1dx\,f(x) = \text{const} + \frac{2}{\gamma x_1}[1+O(x_1)].
$$
Thus, the leading term in the sum is $ 6/\gamma x_1 \approx
12N/\gamma\pi$. So the leading term in $\langle L^2\rangle_T -
\langle L\rangle_T^2 $ is $ (12k_BT/\pi^2\kappa^2)N$, and we obtain
the final result valid for large $N$
\begin{equation}
\frac{\Delta L}{\langle L\rangle_T} \approx
\frac{\sqrt{12k_BT}}{\pi\kappa\xi}\frac{1}{\sqrt{N}}.
\end{equation}
Thus, the dispersion of $L$ is relatively small for large $N$. In
the sense explained in Section 4, the length is a classical property
of our model body.

Clearly, this length and its dispersion are intrinsic properties of our model
body because the conditions that define them are of the intrinsic character.
We have specified the structure in terms of a Hamiltonian, and we have
asked about the average values of some quantity under the assumptions
that the average energy has some value and that the state is the most
probable state satisfying such a condition (maximal entropy). The state is
then the Gibbs state with certain temperature. The probability to find the
body in such a state is very close to unity under the given conditions,
hence the state is a good approximation to what can be observed.

Some comment is in order. First, the thermodynamical equilibrium can
settle down starting from an arbitrary state only if some weak but
non-zero interaction exists between the phonons. Second, the bulk
motion of the chain is decoupled from all other degrees of freedom
and has a character of a closed subsystem with just one degree of
freedom (it is not a macroscopic property, at least of our model). A
different model (including, may be, also other bodies) is needed to
show that any properties of large systems can lead to the
classicality of the bulk motion. Work on this problem is in
progress.

\subsection*{Acknowledgements}
P.H. is indebted to Juerg Gasser and Uwe-Jens Wiese for reading the
manu\-script and suggesting many improvements in the text. Thanks go
to the Institute of Theoretical Physics, Faculty of Mathematics and
Physics of the Charles University, Prague for hospitality and
discussion.
 J.T. gratefully acknowledges partial
support by the Ministry of Education of Czech Republic (projects
MSM6840770039 and LC06002).


\begin{thebibliography}{99}
\bibitem{Griffiths}R.~B.~Griffiths, {\it Consistent Quantum Theory},
Cambridge University Press, Cambridge, UK, 2002.
\bibitem{Nikolic}H.~Nikolic, {\it Quantum Mechanics: Myths and Facts}.
ArXiv: quant-phys/0609163.
\bibitem{Peres}A.~Peres, {\it Quantum Theory: Concepts and Methods},
Kluwer, Dordrecht, 1995.
\bibitem{d'Espagnat}B. d'Espagnat, {\em Veiled Reality}, Addison-Wesley,
Reading, 1995.
\bibitem{Isham}C.~J.~Isham, {\it Lectures on Quantum Theory. Mathematical
and Structural Foundations}, Imperial College Press, London 1995.
\bibitem{Piron}C.~Piron, {\it Foundations of Quantum Physics}, Benjamin,
Reading 1976; C.~Piron, Found.\ Phys.\ {\bf 2} (1972) 287.
\bibitem{BvN}G.~Birkhoff and J.~von~Neumann, {\it The Logic of Quantum
Mechanics}, Ann. of Math. {\bf 37} (1936) 823.
\bibitem{ludwig}G. Ludwig, {\it Foundations of Quantum Mechanics},
Springer, Berlin, 1983.
\bibitem{kraus}K. Kraus, {\it States, Effects, and Operations}, Springer
Lecture Notes in Physics 190, Berlin, 1983.
\bibitem{Exner}F.~Exner, {\it Vorlesungen \"{u}ber die physikalischen
Grundlagen der Naturwissenschaften}, Deuticke, Leipzig, 1922.
\bibitem{Born}M.~Born, Phys.~Bl\"{a}tter {\bf 11} (1955) 49.
\bibitem{thirring}W. Thirring, {\it Lehrbuch der Mathematischen Physik.
4. Quantenmechanik grosser Systeme}, Springer, Berlin, 1980.
\bibitem{bell1}J. S. Bell, {\it The theory of local beables} in {\it
Speakable and Unspeakable in Quantum Mechanics}, Cambridge
University Press, Cambridge (England), 1987.
\bibitem{Leggett}A.~J.~Leggett and A.~Garg, Phys.\ Rev.\ Lett.\ {\bf 54}
(1985) 857; A.~J.~Leggett, J. Phys.: Cond.\ Mat.\ {\bf 14} (2002) R415.
\bibitem{JvN}J.~von~Neumann, {\it Mathematical Foundation of Quantum
Mechanics}, Princeton University Press, Princeton NJ, 1983.
\bibitem{Jauch}J.~M.~Jauch, Helv. Phys. Acta {\bf 37} (1964) 293.
\bibitem{TH}J.~Tolar and P.~H\'aj\'{\i}\v{c}ek,
Phys. Letters A {\bf 353} (2006) 19.
\bibitem{bell2}J. S. Bell, {\it Against `measurement`} in {\it Sixty Two
Years of Uncertainty}, A. I. Miller (Ed.), Plenum, New York, 1990.
\bibitem{zurek}W.~H.~Zurek, Rev.\ Mod.\ Phys.,{\bf 75} (2003) 715.
\bibitem{zeh}D.~Giulini, E.~Joos, C.~Kiefer, J.~Kupsch, I.-O.~Stamatescu,
H.~D.~Zeh, {\it Decoherence and the Appearance of Classical World in
Quantum Theory}, Springer, Berlin, 1996.
\bibitem{Hepp}K.~Hepp, Helvetica Phys.~Acta, {\bf 45} (1972) 237.
\bibitem{Bell3}J.~S.~Bell, Helv. Phys.~Acta, {\bf 48} (1975) 93.
\bibitem{Bona}P.~B\'ona, Acta Phys.~Slov., {\bf 23} (1973) 149, {\bf 25}
(1975) 3, {\bf 27} (1977) 101.
\bibitem{Sewell}G.~L.~Sewell, {\it Quantum Mechanics and its Emergent
Macrophysics}, Princeton University Press, Princeton, 2002.
\bibitem{poulin} D.~Poulin, Phys.\ Rev.\ A {\bf 71} (2005) 022102.
\bibitem{Kofler}J.~Kofler and \v{C}.~Brukner, Phys.\ Rev.\ Lett. {\bf 99}
(2007) 180403.
\bibitem{Kittel}C.~Kittel,
 {\it Introduction to Solid State Physics}, Wiley, New York 1976.
\bibitem{Rutherford} D.~E.~Rutherford, Proc.\ Roy.\ Soc.\ (Edinburgh),
Ser.~A, {\bf 62} (1947), 229; {\bf 63} (1951), 232.

\end{thebibliography}
\end{document}